\documentclass[aps,prd,12pt,onecolumn,tightenlines,showkeys,nofootinbib,amsmath,amssymb]{revtex4-2}
\usepackage{graphicx}
\usepackage{booktabs}
\usepackage{bm}
\usepackage[hidelinks]{hyperref}
\usepackage{xcolor}
\usepackage{comment}

\newcommand{\phiG}{\varphi}
\newcommand{\epsql}{\epsilon_{ql}}
\newcommand{\epsC}{\epsilon_C}
\newcommand{\epsFN}{\epsilon_{\scriptscriptstyle\mathrm{FN}}}

\newcommand{\arcosh}{\operatorname{arcosh}}

\newcommand{\cF}[1]{c_{#1{\scriptscriptstyle F}}}

\begin{document}

\title{Fine structure of fermion mixing and quark-mass hierarchies}

\author{Jacek Ciborowski}
\affiliation{Faculty of Physics, University of Warsaw, L. Pasteura 5, PL-02-093 Warsaw, Poland}

\date{30 October 2026}

\begin{abstract}
We investigate a semi-empirical flavour framework based on two linked hypotheses: a golden/pentagonal organisation of fermion mixing and the recurrence of its hierarchy maps in the quark-mass spectrum. The two large lepton angles are specified by the fifth-order Chebyshev conditions $T_5(\theta_{12}^{\ell})=0$ and $T_5(\theta_{23}^{\ell})=-1$, with angular arguments in radian measure, while the smaller reactor angle follows the approximate bilinear relation $\theta_{13}^{\ell}\simeq\theta_{12}^{\ell}\theta_{23}^{\ell}/\pi$. An alternative golden-partition construction yields nearly identical large-angle predictions and an exact circular--hyperbolic duality.

The golden ratio $\phiG$ also enters two distinct geometrical prescriptions for the Cabibbo angle, while the heptadecagonal seed $\alpha_C=\pi/17$ governs the suppressed mixing hierarchy. The near-equality $\tan\alpha_C\simeq\cos54^\circ/\pi$ further connects the two geometries. The resulting quark hierarchy obeys the quadratic map $\theta_{13}^q=\sqrt{\theta_C^2+(\theta_{23}^q)^2}-\theta_C\simeq(\theta_{23}^q)^2/(2\theta_C)$. Transferring the quadratic and bilinear maps to quark masses yields the crossed relations $m_d\simeq m_c^2/(2m_t)$ and $m_u\simeq  m_b m_s/(m_F/\phiG)$, where $m_F=G_F^{-1/2}$. The complementary triples $(t,c,d)$ and $(b,s,u)$ reproduce both mass targets within $1\%$ for the adopted reference values. Gatto--Sartori--Tonin-type correspondences and Froggatt--Nielsen scaling provide compatible phenomenological interpretations.

These results suggest a common golden/geometrical organising structure for flavour mixing and mass hierarchies, whose underlying dynamical origin remains to be established.
\end{abstract}

\maketitle

\section{Introduction}
\label{sec:Introduction}

The Standard Model (SM) accommodates fermion masses and flavour mixing through Yukawa matrices, but does not explain the observed numerical values or the strong hierarchies among them.  The CKM and PMNS matrices provide an economical parametrization of quark and lepton mixing, while the Froggatt--Nielsen (FN) mechanism offers a broad framework in which hierarchical Yukawa couplings arise as powers of a small symmetry-breaking parameter~\cite{cite:FN}.  The Gatto--Sartori--Tonin (GST) relation provides a complementary connection between a mixing angle and a fermion-mass ratio~\cite{cite:GST}.  In both approaches, however, order-one coefficients or texture assumptions remain essential.

In earlier work we proposed a ``Bi-Constructible'' phenomenology in which selected weak and flavour mixing angles are organized by the Euclidean geometry of regular pentagons and heptadecagons, associated with the Fermat primes $5$ and $17$~\cite{cite:Ciborowski2025}.  The present work turns that observation into two explicit tests.  The first is a parameter-free mixing hypothesis, with $P5$ and $G$ as two discrete variants for the two large lepton angles and $P17$ supplying the suppression hierarchy.  The second is a cross-sector transfer hypothesis: the two hierarchy maps identified in mixing are tested for recurrence in the six quark masses.  GST, FN scaling, RG evolution and symmetry considerations are then used as secondary interpretations and consistency checks rather than as additional primary hypotheses.

The guiding philosophy is deliberately semi-empirical. Throughout the analysis, we distinguish between (i) exact mathematical identities, (ii) experimentally testable numerical correspondences, and (iii) possible theoretical interpretations. In particular, numerical agreement alone is not regarded as evidence for a fundamental physical law. Rather, the principal motivation is the apparent recurrence of a small set of mathematical structures---fivefold and seventeenfold angular relations and two induced-hierarchy maps---across otherwise distinct flavour observables, supplemented by connections to square-root Gatto--Sartori--Tonin (GST) relations and Froggatt--Nielsen (FN) scaling.

Empirical correspondences are assessed against the relevant experimental uncertainties, exact results are established only within their explicitly stated mathematical assumptions, and possible underlying mechanisms are treated as model-building hypotheses rather than derived physical consequences. Unless stated otherwise, propagated numerical uncertainties use first-order independent-input error propagation; input correlations, scheme-matching uncertainties and theoretical errors of the empirical ansatzes are not included. Accordingly, the mathematical form of a displayed equation does not, by itself, establish its physical status. Each relation is therefore identified in the accompanying discussion as an exact identity, a consequence of stated assumptions, an empirical ansatz or correspondence, or an exploratory observation.

To make the logical dependencies explicit, we organise the analysis around two primary modules. The first, $H_{\rm mix}^{X}$, with $X=P5$ or $G$, postulates that the two large lepton mixing angles are fixed by one of two discrete pentagonal/golden prescriptions, while the hierarchy of the smaller mixing angles is governed by the heptadecagonal seed $\alpha_C\equiv\pi/17$. In the quark sector, the radius $R_{17}=1/\varphi$ is introduced as an explicit cross-sector normalisation ansatz. Once the branch $X$ is selected, no continuous parameter is adjusted to reproduce the mixing-angle magnitudes.

The second module, $H_{\rm tr}$, introduces a cross-sector universality hypothesis: the quadratic map $\mathcal Q(a,b)=\sqrt{a^2+b^2}-a$ and the bilinear map $\mathcal B(a,b;S)=ab/S$, identified in the mixing construction, are conjectured to recur in the quark-mass spectrum. The bilinear mass scale $S=m_F/\varphi$, where $m_F$ denotes the Fermi scale and $\varphi$ the golden ratio, is held fixed in the present comparison; its identification was motivated by the same quark-mass data. The resulting mass relations are therefore not deductions from the mixing angles, but constitute an additional phenomenological hypothesis concerning the recurrence of the same hierarchy maps in the mass sector.

The distinction between goodness of description and discovery significance is essential. Since the existing experimental data contributed to the identification of the proposed patterns, the statistical diagnostics presented below cannot be interpreted as look--elsewhere-corrected discovery significances. Rather, they quantify the agreement of the now-fixed hypotheses with the adopted observables. Future independent measurements, evaluated without readjusting the hypotheses or their defining assumptions, can therefore provide genuine prospective tests of the proposed structure.

A related methodological point concerns the look--elsewhere problem inherent in any semi-empirical search for numerical structures. Our aim is not to test all possible functions or integer decompositions, but to examine a low-complexity class of hypotheses motivated by pentagonal/heptadecagonal geometry and the associated golden-ratio algebra. The primitive orders $5$ and $17$ are identified in the mixing sector and are not reselected for individual observables. By contrast, the large mass-sector indices provide continuous parametrisations of the reference hierarchies; nearby integer assignments are retained only as exploratory arithmetic observations, not as requirements of the principal mass relations (Sec.~\ref{subsec:425}).

Throughout the analysis, mixing angles entering algebraic equations are expressed in radians; degree values are provided only for readability. The golden ratio $\phiG=(1+\sqrt5)/2$ is the positive root of $\phiG^2-\phiG-1=0$. Its recurrence plays an organising role, although its various appearances have different logical status. In the $P5$ scheme, the golden ratio follows directly from the Euclidean geometry of the regular pentagon, notably through $\theta_{23}^{\ell(P5)}=\phiG/2$ and the corresponding radical expression for $\theta_{12}^{\ell(P5)}$. The reciprocal $G$ roots and the two symmetry-based predictions, GR1 and GR2, represent related manifestations of the same golden algebra rather than statistically independent recurrences. The radius $1/\phiG$ employed in the $P17$ construction is an explicit cross-sector normalisation ansatz inherited from the lepton/golden structure; its agreement with the Cabibbo angle constitutes the empirical result. In contrast, the mass relation involving $m_F/\phiG$, where $m_F\equiv G_F^{-1/2}$ and $G_F=1.16637859(59)\cdot 10^{-5}\,\mathrm{GeV}^{-2}$ is the Fermi constant~\cite{cite:FermiConstantUpdate2026}, and the top--$G$ correspondence relate independently measured quantities from different sectors to values fixed by the golden construction. They therefore constitute additional cross-sector empirical correspondences, albeit identified within the same restricted phenomenological programme.

Using the current value of $G_F$ gives $m_F\simeq292.806~\mathrm{GeV}$~\cite{cite:PDG2026}. We adopt $m_F/\phiG$ as a phenomenological normalisation. Since $m_F=2^{1/4}v$, where $v$ is the conventional electroweak vacuum scale, no independent dynamical significance is attributed to this choice. The $m_F$ convention is retained because it gives particularly simple expressions for the top--$G$ correspondence and the bilinear mass scale.

\section{Lepton mixing: a fivefold spectral structure}
\label{sec:Leptons}

\subsection{Chebyshev conditions and the $P5$ scheme}
\label{subsec:ChebyshevP5}

The two large lepton mixing angles lie close to a node and an
extremum of the fifth Chebyshev polynomial,
$T_5(x)=16x^5-20x^3+5x$. For the adopted PDG~2026
benchmarks one finds
\begin{equation}
T_5\!\left(\theta_{12}^{\ell(\mathrm{exp})}\right)
 =0.003\pm0.080,\qquad
T_5\!\left(\theta_{23}^{\ell(\mathrm{exp})}\right)
 =-0.997\pm0.013.
\label{eq:ChebyshevExperimental}
\end{equation}
The solar input is $\sin^2\theta_{12}^{\ell}=0.307\pm0.012$;
the more recent JUNO determination,
$\sin^2\theta_{12}^{\ell}=0.3036\pm0.0064$
\cite{cite:JUNO2026Update}, gives
$T_5(\theta_{12}^{\ell})=0.026\pm0.043$, consistent with
zero within $0.6\sigma$. For normal ordering, the PDG
atmospheric input is $\sin^2\theta_{23}^{\ell}=0.533\pm0.018$,
while IceCube DeepCore finds $0.54^{+0.04}_{-0.03}$
\cite{cite:PDG2026NuMix,cite:IceCube2025}. Both central values
lie in the upper octant, although the atmospheric likelihood
retains octant ambiguities and non-Gaussian features
\cite{cite:PDGPMNSParam}.

The atmospheric uncertainty in Eq.~\eqref{eq:ChebyshevExperimental}
is only a local linear estimate, not a Gaussian confidence interval
for the transformed quantity. At the $P5$ point,
$T_5'(\theta_{23}^{\ell(P5)})=0$, so $T_5+1$ varies quadratically
with departures from the predicted angle. The transformed likelihood
is consequently asymmetric and bounded below by $T_5=-1$ in the
relevant interval. Quantitative discrimination should therefore use
the original angle likelihood; proximity to the polynomial minimum
is not an additional, independent precision test.

Motivated by this agreement, we impose the idealised conditions
\begin{equation}
T_5(\theta_{12}^{\ell})=0,\qquad
T_5(\theta_{23}^{\ell})=-1.
\label{eq:ChebyshevIdeal}
\end{equation}
The polynomial acts on the mixing-angle values themselves,
not on their cosines. Introducing auxiliary phases by
$\theta_{12}^{\ell}=\cos\alpha$ and
$\theta_{23}^{\ell}=\cos\beta$, the identity
$T_5(\cos\gamma)=\cos(5\gamma)$ turns
Eq.~\eqref{eq:ChebyshevIdeal} into the node--extremum
conditions $\cos(5\alpha)=0$ and $\cos(5\beta)=-1$.
In the positive nontrivial range, the first permits
$\alpha=18^\circ$ or $54^\circ$, while the second selects
$\beta=36^\circ$ and hence
$\theta_{23}^{\ell}=\cos36^\circ=\phiG/2$.
The latter is a double root of $T_5(x)+1$, corresponding
to an extremum of the fifth harmonic.

Requiring the two large angles to form a unit-normalised
pair,
\begin{equation}
(\theta_{12}^{\ell})^2+(\theta_{23}^{\ell})^2=1,
\label{eq:P5UnitNorm}
\end{equation}
then selects $\alpha=\alpha_{P5}\equiv54^\circ$ and uniquely fixes
the physical positive solution
\begin{align}
\theta_{12}^{\ell(P5)}
 &=\cos54^\circ
 =\sqrt{1-\left(\frac{\phiG}{2}\right)^2}
 \simeq0.587785\simeq33.6776^\circ,
\label{eq:Theta12P5}\\
\theta_{23}^{\ell(P5)}
 &=\sin54^\circ=\frac{\phiG}{2}
 \simeq0.809017\simeq46.3533^\circ.
\label{eq:Theta23P5}
\end{align}
Thus, once the fifth-order conditions and the common
normalisation are assumed, the pentagonal direction is
selected from discrete possibilities rather than introduced
as an adjustable angular parameter.

The node--extremum description is an auxiliary spectral
interpretation, not a derivation of a physical standing wave
in flavour space. Moreover, $n=5$ is the lowest order
reproducing these \emph{particular} spectral phases: all
higher odd multiples also satisfy the same conditions through
$T_{5(2k+1)}(x)=T_{2k+1}(T_5(x))$.
The selection of the primitive order $5$, associated with
pentagonal geometry, remains an empirical ingredient of
the framework.

\subsection{The alternative golden-partition scheme}
\label{subsec:GoldenScheme}

The golden partition remains an ansatz rather than a dynamical derivation: the present construction does not explain why flavour dynamics should select the golden reciprocal pair, as shown below.  It is, however, not an arbitrary parametrization introduced separately for the two measured angles.  The golden ratio is intrinsic to the pentagonal $P5$ geometry, and once the reciprocal pair is adopted, the compact solar branch and the hyperbolic atmospheric branch follow together from the same algebraic input.  In this sense the $G$ construction forms the second side of the $P5/G$ dual description rather than an unrelated numerical fit.

The following  set of equations describes the golden partition of the full angle: 
$\beta / \alpha = \alpha / ( \alpha+\beta)$   and   $\alpha + \beta = 1$,
where the angles $\alpha$ and $\beta$ were  normalized to $2\pi$.  The above set   has  the following two pairs of solutions, $\alpha_{\pm}$ and $\beta_{\pm}$:  $ \alpha_{+} =-\varphi$,   $\beta_{+} =\varphi^2$ 
and $ \alpha_{-} =\varphi -1$,  $\beta_{-} =\varphi^{-2}$.

Since $0<\beta_-<1$, the first root lies in the range of a real cosine and defines an angle approximately equal the $P5$ solar angle through
$\cos(2\theta_{12}^{\ell(G)})=\beta_-$.  Applying the analogous equation to the reciprocal root,
$\cos(2\chi)=\beta_+$, gives no real solution because $\beta_+>1$.  Writing instead $\chi=i\eta$ analytically continues the circular function according to $\cos(2i\eta)=\cosh(2\eta)$.  The positive real continuation therefore defines the angle approximately equal the $P5$ atmospheric angle through $\cosh(2\theta_{23}^{\ell(G)})=\beta_+$.  Thus the cosine and hyperbolic-cosine forms are not two independently chosen ansatzes, but arise from the two branches of the same reciprocal golden construction:
\begin{equation}
\cos\left(2\theta_{12}^{\ell(G)}\right)=\frac{1}{\phiG^2},
\qquad
\cosh\left(2\theta_{23}^{\ell(G)}\right)=\phiG^2.
\label{eq:GoldenConditions}
\end{equation}
The same circular--hyperbolic continuation is built directly into the Chebyshev structure.  Besides
$T_n(\cos\alpha)=\cos(n\alpha)$, one has for $x\geq 1$: 
$T_n(x)=\cosh\!\big(n\,\arcosh(x)\big)$ or equivalently $T_n(\cosh\eta)=\cosh(n\eta)$. 
This identity suggests a possible auxiliary algebraic representation of the $G$ construction.  For a unimodular $2\times2$ matrix with half-trace $x=\mathrm{Tr}\,M/2$, the reciprocal roots $\phiG^{-2}<1$ and $\phiG^{2}>1$ lie on opposite sides of the trace boundary $|x|=1$: the first belongs to the circular branch, whereas the second belongs to its hyperbolic continuation.  This trace classification is used only to organize the two mathematical branches; no noncompact physical flavour degree of freedom is assumed.  The observation therefore suggests a common underlying algebraic construction for the solar and atmospheric $G$ relations, rather than identifying a microscopic flavour symmetry.
Equivalently,
\begin{equation}
\theta_{12}^{\ell(G)} =\frac12\arccos\left(\phiG^{-2}\right)
\simeq33.7722^\circ,  \qquad \theta_{23}^{\ell(G)} =\frac12\arcosh\left(\phiG^2\right)\simeq46.3214^\circ.
\label{eq:Theta12G23G}
\end{equation}

Both the $P5$ and $G$ constructions therefore predict $\theta_{23}^{\ell}>45^\circ$, placing the atmospheric angle in the second (upper) octant, only about $1.3^\circ$ above maximal mixing.  This is a qualitative prediction common to the two near-degenerate branches: a future determination that robustly selects the first octant would directly disfavor both constructions.

The circular--hyperbolic correspondence is expressed by
$\tan^2\theta_{12}^{\ell(G)}=(1-\beta_-)/(1+\beta_-)
=(\beta_+-1)/(\beta_++1)=\tanh^2\theta_{23}^{\ell(G)}$, so that
\begin{equation}
\tan\theta_{12}^{\ell(G)}
=
\tanh\theta_{23}^{\ell(G)}
=
5^{-1/4}.
\label{eq:CircularHyperbolicDuality}
\end{equation}
They also imply
\begin{equation}
\sin^2\theta_{12}^{\ell(G)}=\frac{1}{2\phiG}\simeq 0.3090.
\label{eq:GoldenSolarProbability}
\end{equation}
A further exact relation connects Eq.~\eqref{eq:GoldenSolarProbability} to the two standard golden-ratio solar-mixing ansatzes usually denoted GR1 and GR2~\cite{cite:GoldenGR2,cite:GoldenA5}.  In the GR1 scheme,
$\tan\theta_{12}^{(\mathrm{GR1})}=1/ \phiG$ and $\sin^2\theta_{12}^{(\mathrm{GR1})}=1 / (\phiG\sqrt5)\simeq 0.2764$~$\big(\theta_{12}^{(\mathrm{GR1})}\simeq 31.7175^\circ\big)$, 
whereas in the GR2 scheme,
$\cos\theta_{12}^{(\mathrm{GR2})}=\phiG / 2$ and $ \sin^2\theta_{12}^{(\mathrm{GR2})}=\sqrt5/ (4\phiG )\simeq 0.3455$~$\big(\theta_{12}^{(\mathrm{GR2})}\simeq 36^\circ\big)$.
It follows exactly that
\begin{equation}
\sin^2\theta_{12}^{\ell(G)}
=
\sqrt{
\sin^2\theta_{12}^{(\mathrm{GR1})}
\sin^2\theta_{12}^{(\mathrm{GR2})}
}
=
\frac{1}{2\phiG}
.
\label{eq:GoldenGeometricMean}
\end{equation}
Thus the $G$ value is the exact geometric mean of the two previously studied golden-ratio mixing probabilities.  In this precise sense the $G$ prescription need not be viewed as a third unrelated golden ansatz: it interpolates multiplicatively between the GR1 and GR2 structures.  We stress that the geometric-mean relation applies to the probabilities $\sin^2\theta_{12}$ rather than to the angles themselves.  At the level of the angles one nevertheless has the simple ordering
\begin{equation}
\theta_{12}^{(\mathrm{GR1})}
<\theta_{12}^{\ell(P5)}
\lesssim\theta_{12}^{\ell(G)}
<\theta_{12}^{(\mathrm{GR2})}.
\label{eq:GoldenSolarOrdering}
\end{equation}

The two values $P5/G$  differ by only $\Delta\theta_{12}^{\ell}\simeq0.1^\circ$ and
$\Delta\theta_{23}^{\ell}\simeq-0.03^\circ$. Moreover, the $G$ pair has Euclidean norm
$R^{(G)}\simeq1.000522$ and polar direction $\alpha^{(G)}\simeq53.9048^\circ$, remarkably close
to the unit $P5$ vector at $54^\circ$. We therefore regard them as a near-degenerate pair
of mathematically distinct geometrical prescriptions, presently indistinguishable experimentally.

A microscopic flavour theory would have to explain why the golden ratio appears to govern the two large lepton mixing angles and whether the mathematically distinct $P5$ and $G$ prescriptions,  originate from a common underlying mechanism.

\subsection{The reactor angle and the heptadecagonal seed}
\label{subsec:LeptonP17}

The PDG~2026 reactor angle, $\sin^2\theta_{13}^{\ell}=0.0217\pm0.0007$, i.e., 
$\theta_{13}^{\ell(\mathrm{exp})}=(8.47\pm0.14)^\circ$
\cite{cite:PDG2026NuMix}, is substantially smaller than
the other two lepton mixing angles. Together with the
PDG~2026 atmospheric angle (normal ordering), it gives
the experimental suppression ratio
\begin{equation}
\epsilon_{\ell}^{(\mathrm{exp})}
\equiv\frac{\theta_{13}^{\ell(\mathrm{exp})}}
{\theta_{23}^{\ell(\mathrm{exp})}}
=0.1806\pm0.0049.
\label{eq:EpsilonLeptonExperimental}
\end{equation}
For comparison, NuFIT~6.1 quotes
$\theta_{13}^{\ell}=(8.62\pm0.11)^\circ$ for normal
ordering~\cite{cite:NuFIT61}, approximately $1.8\%$
higher than the PDG value. At the same atmospheric
reference angle, this would increase the suppression
ratio to about $0.1838$.

The PDG suppression ratio corresponds to an effective
polygonal order
\[
n_{\ell}^{(\mathrm{eff})}
=\frac{\pi}{\arctan\epsilon_{\ell}^{(\mathrm{exp})}}
=17.58\pm0.47.
\]
Although the lepton data alone do not select a unique
integer, the analogous quark hierarchy discussed below
gives $n_q^{(\mathrm{eff})}=17.22\pm0.31$, favouring
the common order $17$. This is also the next nontrivial
Fermat prime after $5$, associated with a constructible
regular polygon. We therefore adopt the heptadecagonal
angle $\alpha_C=\pi/17\simeq10.5882^\circ$
and the corresponding geometrical suppression parameter
$\epsC\equiv\tan\alpha_C\simeq0.186932$.
The latter is consistent with the experimental lepton
ratio within approximately $1.3\sigma$.
The $P5$ lepton hierarchy then predicts
\begin{equation}
\theta_{13}^{\ell(P5)}
=\theta_{23}^{\ell(P5)}\epsC
\simeq8.6649^\circ.
\label{eq:Theta13P5}
\end{equation}

Several close, but not exact, numerical relations connect
the pentagonal and heptadecagonal geometries:
\begin{align}
\cos\alpha_{P5}
&\simeq\pi\tan\alpha_C,
\label{eq:P5P17Bridge1}\\
\cos\alpha_{P5}+\cos\alpha_C
&\simeq\frac{\pi}{2},
\label{eq:P5P17Bridge2}\\
\frac{2\pi}{17\phiG}\cos\alpha_C
&\simeq\frac{\cos\alpha_{P5}}{\phiG^2}.
\label{eq:P5P17Bridge3}
\end{align}
Their relative discrepancies are approximately
$8.8\times10^{-4}$, $2.4\times10^{-5}$, and
$9.6\times10^{-5}$, respectively. These are
arithmetic near-relations between the two geometries,
not independent experimental confirmations.

Using Eq.~\eqref{eq:P5P17Bridge1}, the reactor-angle relation~\eqref{eq:Theta13P5} takes the approximate bilinear form
\begin{equation}
\theta_{13}^{\ell}
\simeq\frac{\theta_{12}^{\ell}\theta_{23}^{\ell}}{\pi},
\label{eq:LeptonBilinear}
\end{equation}
giving $8.6726^\circ$ for the exact $P5$ inputs.
Thus the factor $\pi$ follows from the approximate
$P5$--$P17$ geometrical connection rather than
being introduced as an adjustable scale.
The possible recurrence of this bilinear structure
in the quark-mass sector is examined below.

\subsection{PMNS-modulus formulation}
\label{subsec:PMNSModuli}

The lepton mixing relations can also be expressed through
rephasing-invariant PMNS moduli. For a unitary three-neutrino
mixing matrix,
\begin{equation}
\theta_{12}^{\ell}=\arctan\frac{|U_{e2}|}{|U_{e1}|},\quad
\theta_{23}^{\ell}=\arctan\frac{|U_{\mu3}|}{|U_{\tau3}|},\quad
\theta_{13}^{\ell}=\arcsin|U_{e3}|.
\label{eq:PMNSAnglesFromModuli}
\end{equation}
Consequently, the $P17$ suppression relation becomes
\begin{equation}
|U_{e3}|=\sin\!\left[
\epsilon_C\,
\arctan\frac{|U_{\mu3}|}{|U_{\tau3}|}
\right].
\label{eq:PMNSP17HierarchySine}
\end{equation}
The $P5$ Chebyshev conditions can likewise be expressed
through the modulus ratios above. These relations require
no assumption about the CP phase and provide alternative
rephasing-invariant formulations of the same mixing
hypotheses, rather than independent experimental constraints.

\subsection{Geometrical mixing patterns and neutrino anarchy}
\label{subsec:Anarchy}

Neutrino anarchy offers a statistical description in which
the invariant Haar measure determines distributions of PMNS
parameters without selecting particular mixing-angle
values~\cite{cite:Anarchy,cite:HaarAnarchy}.
The present framework instead asks whether these angles
participate in a common structure extending across flavour
sectors.

The significance of the $P5/G$ prescriptions lies not merely
in their agreement with the two large lepton angles, but in
their integration into the wider pattern of golden-ratio
normalisations and a common $P17$ suppression hierarchy.
An anarchical description of lepton mixing would not, by
itself, account for these particular cross-sector connections:
they would remain accidental correspondences or require
additional structure. Their persistence under increasingly
precise measurements would therefore strengthen the case
for a common flavour origin. This structural motivation
does not yet establish a statistical preference over anarchy;
a quantitative comparison would have to account for
experimental uncertainties and the a posteriori identification
of the geometrical relations.

\section{Quark mixing and the $P17$ geometry}
\label{sec:QuarkMixing}

\subsection{Experimental reference values}
\label{subsec:CKMInputs}

The 2026 PDG global fit with three-generation unitarity gives~\cite{cite:PDG2026CKM}
$\sin\theta_{12}^q=0.22517\pm0.00068$, $\sin\theta_{23}^q=0.04189^{+0.00081}_{-0.00069}$ and 
$\sin\theta_{13}^q=0.003763^{+0.000088}_{-0.000083}$. These correspond to approximately
$\theta_{12}^q=(13.01\pm0.04)^\circ$,
$\theta_{23}^q=(2.40^{+0.05}_{-0.04})^\circ$, and
$\theta_{13}^q=(0.216\pm0.005)^\circ$.
For tests of a flavour hierarchy it is also useful to quote separately determined elements before imposing the global CKM fit.  Representative values used below are $|V_{us}|=0.22431\pm0.00085$, $|V_{cb}|\simeq0.0411\pm0.0012$, and $|V_{ub}|\simeq0.00382\pm0.00020$~\cite{cite:PDG2026CKM,cite:PDG2026Vus}.  Accordingly, we use the global-fit values when testing relations formulated explicitly in the standard CKM angles, while separately determined CKM moduli are used for complementary, rephasing-invariant consistency checks.

The precision of the global-fit angles should not, however, be confused with complete consistency among the underlying determinations.  In the kaon sector, the current semileptonic determination gives $|V_{us}|=0.22330\pm 0.00057$ from $K_{\ell3}$ decays, whereas leptonic $K_{\mu2}/\pi_{\mu2}$ decays give $|V_{us}|/|V_{ud}|=0.23108\pm 0.00051$; the PDG kaon average $|V_{us}|=0.22431\pm 0.00085$ therefore carries a scale factor $S=2.5$ to account for the discrepancy~\cite{cite:PDG2026Vus}.  The heavy-flavour sector shows a similar cautionary example: current inclusive and exclusive determinations give $|V_{cb}|=(42.0\pm0.5)\times10^{-3}$ and $(39.5\pm0.5)\times10^{-3}$, respectively, whose average has $p(\chi^2)<1\%$~\cite{cite:PDG2026VcbVub}.  By contrast, the present inclusive and exclusive values of $|V_{ub}|$, $4.06\times10^{-3}$ and $3.75\times10^{-3}$ with their quoted experimental and theoretical uncertainties, are compatible within about one standard deviation~\cite{cite:PDG2026VcbVub}.  These spreads motivate separate comparisons with the unitary global fit and the directly determined CKM moduli; they do not remove the quantitative preference of the adopted global fit for one Cabibbo normalisation (Sec.~\ref{subsec:GoldenP17}). Further convergence of the underlying experimental and lattice inputs is needed to establish the robustness of that preference.

We do not attempt to derive CKM unitarity from the geometrical construction.  Once the proposed mixing angles, supplemented where needed by the experimentally fitted CKM phase, are inserted into the standard three-generation parametrization, unitarity follows identically from that parametrization; the nontrivial phenomenological test is instead agreement with independently determined CKM observables.  No geometrical prediction of the CP phase is imposed.

For the phenomenological hierarchy analysis we also introduce the quark-sector ratio
\begin{equation}
\epsilon_q^{(\mathrm{exp})}
\equiv\frac{\theta_{23}^q}{\theta_C}
=0.1845\pm0.0034,
\label{eq:EpsilonQuarkExperimental}
\end{equation}
where the asymmetric $\theta_{23}^q$ error is symmetrised for this comparison.
The above and the  lepton determination Eq.~\eqref{eq:EpsilonLeptonExperimental}   are consistent and 
their inverse-variance weighted mean, adopted in the FN coefficient analysis, is
\begin{equation}
\epsql=0.1833\pm0.0028.
\label{eq:EpsilonAdopted}
\end{equation}

\subsection{Two golden-ratio constructions of the Cabibbo angle}
\label{subsec:GoldenP17}

The Cabibbo angle admits two distinct  golden-ratio prescriptions, each parameter-free once its cross-sector normalisation is specified.
The first assigns to it the smaller golden fraction of the
pentagonal solar angle:
\begin{equation}
\theta_C^{(\mathrm{sol})}
=\frac{\theta_{12}^{\ell(P5)}}{\phiG^2}
=\frac{\cos54^\circ}{\phiG^2}
\simeq12.8637^\circ.
\label{eq:SolarCabibboRelation}
\end{equation}
The associated right-triangle hypotenuse is
$R_C^{(\mathrm{sol})}=\theta_C^{(\mathrm{sol})}/\cos\alpha_C$.
The sum of seventeen such equal angular segments satisfies $17R_C^{(\mathrm{sol})}\simeq2\pi/\phiG$, with a relative
residual of $\simeq 10^{-4}$. Thus the smaller golden fraction of the solar angle and the larger golden fraction of a full
angle, complementary through $\phiG^{-2}+\phiG^{-1}=1$, enter the same construction.  Using the $G$ solar value instead gives
$\theta_{12}^{\ell(G)}/\phiG^2\simeq12.8998^\circ$.

The second prescription identifies the Cabibbo angle with a side  of a regular heptadecagon inscribed in a circle of radius
$R_{17}=1/\phiG$:
\begin{equation}
\theta_C^{(P17)}
=\frac{2}{\phiG}\sin\alpha_C
\simeq13.0134^\circ.
\label{eq:CabibboChord}
\end{equation}
Here the golden radius is an explicit cross-sector normalisation ansatz, not a consequence of heptadecagonal geometry.
The distinction is therefore between transferring the solar angular scale through a golden partition and imposing a golden
radius directly.

Both prescriptions share the same subsequent hierarchy. For either Cabibbo normalisation, the right triangle gives
$\theta_{23}^q=\theta_C\tan\alpha_C$. Identifying the excess of its hypotenuse over $\theta_C$ with the smallest quark
angle yields
\begin{equation}
\theta_{13}^q
=\sqrt{\theta_C^2+(\theta_{23}^q)^2}-\theta_C
=\theta_{23}^q\tan\frac{\alpha_C}{2}
\simeq\frac{(\theta_{23}^q)^2}{2\theta_C},
\qquad(\theta_{23}^q\ll\theta_C).
\label{eq:QuarkPythagorean}
\end{equation}
The last expression is termed \emph{seesaw-like} only because of its algebraic form. The solar-derived $P5$ normalisation
gives $(\theta_{23}^q,\theta_{13}^q) \simeq(2.4046^\circ,0.2228^\circ)$, whereas the golden-radius
chord gives $(2.4326^\circ,0.2254^\circ)$.

For the adopted unitary global fit, the solar-derived $P5$ prescription
predicts $\sin\theta_C^{(\mathrm{sol})}=0.222633$, about $3.7\sigma$
below $0.22517\pm0.00068$, whereas the chord prescription differs
by only about $0.01\sigma$~\cite{cite:PDG2026CKM}.
Against the separately determined kaon average
$|V_{us}|=0.22431\pm0.00085$~\cite{cite:PDG2026Vus},
the corresponding predictions lie about $2.0\sigma$ below and
$1.0\sigma$ above the central value, respectively, using
$|V_{us}|=\sin\theta_C\cos\theta_{13}^q$.
These are indicative single-observable comparisons, not a combined
fit. The adopted global fit therefore disfavours the solar-derived
Cabibbo normalisation, although that prescription is closer for the
two smaller quark angles. Neither normalisation is derived here from
flavour dynamics. The spread among direct determinations motivates
retaining the solar-derived prescription as an exploratory alternative,
not as one equally supported by the chosen global fit. More precise
and mutually consistent measurements can test whether the preference
persists, or reject both prescriptions. For definiteness, the
Wolfenstein and CKM-modulus benchmarks below retain the chord
normalisation of Eq.~\eqref{eq:CabibboChord}.

\subsection{Wolfenstein and CKM-modulus reformulation}
\label{subsec:WolfensteinP17}

The principal quark hierarchy of the  geometrical construction can also be expressed in the Wolfenstein language  directly through rephasing-invariant CKM moduli.  In the standard exact definition of the leading Wolfenstein parameters one has $\lambda=\sin\theta_{12}^q$ and $ A\lambda^2=\sin\theta_{23}^q$.
Since $|V_{us}|=s_{12}c_{13}$ and $|V_{cb}|=s_{23}c_{13}$, their ratio is exactly
$\epsilon_W\equiv|V_{cb}|/|V_{us}| =\sin\theta_{23}^q/\sin\theta_{12}^q =A\lambda$.
Using the adopted direct determinations quoted above gives $\epsilon_W=0.1832\pm0.0054$,
consistent with Eq.~\eqref{eq:EpsilonAdopted}.
Interpreting this ratio through the polygonal rule $\epsilon_W=\tan(\pi/n_W^{(\mathrm{eff})})$ gives
$n_W^{(\mathrm{eff})}=\pi/\arctan\epsilon_W=17.34\pm0.50$, consistent with the heptadecagonal order $17$.
The next CKM-modulus ratio provides an analogous test of the doubled order.  The direct values give
$ |V_{ub}| / |V_{cb}| =0.09294\pm0.00557$  and hence  $\pi / \arctan(|V_{ub}|/|V_{cb}|) =33.90\pm1.98$, 
which is consistent with the doubled order $34=2\times17$.  
Both ratios are algebraically independent of the CKM phase in the standard parametrization: $|V_{cb}|/|V_{us}|=s_{23}/s_{12}$ and $|V_{ub}|/|V_{cb}|=s_{13}/(s_{23}c_{13})$, where $s_{ij}=\sin\theta_{ij}^{q}$ and $c_{ij}=\cos\theta_{ij}^{q}$~\cite{cite:PDGCKMParam}.  
The substantive comparison concerns the specific orders $17$ and $34$ motivated by the heptadecagonal construction, rather than integers selected separately to match each measured ratio. Both ratios contain the common factor $|V_{cb}|$ and are therefore correlated. The resulting agreements should consequently be regarded as complementary consistency checks rather than statistically independent confirmations.

The $P17$ chord also admits a simple Wolfenstein representation. Equation~\eqref{eq:CabibboChord} predicts
\begin{equation}
\lambda^{(P17)}
=\sin\!\left(\frac{2}{\phiG}\sin\alpha_C\right)
\simeq 0.225179,
\label{eq:LambdaP17}
\end{equation}
in excellent agreement with the global-fit value of $\sin\theta_{12}^q$. Combining this with the geometric hierarchy parameter $\epsilon_C\simeq0.1869$ gives the leading-order Wolfenstein coefficient $A^{(P17)}\simeq\epsilon_C/\lambda^{(P17)}\simeq0.8302$. Using the exact definition $A=\sin\theta_{23}^q/\lambda^2$ instead yields $A^{(P17)}\simeq0.837$, consistent with the global-fit value $A=0.826\pm0.016$. Thus, the otherwise order-one Wolfenstein coefficient $A$ is naturally related to the ratio of the heptadecagonal hierarchy parameter and the Cabibbo/Wolfenstein parameter.

\subsection{Matrix-level hierarchy parameter and a possible FN matching scale}
\label{subsec:FNMatchingScale}

The preceding modulus ratio $|V_{cb}|/|V_{us}|$ is
particularly useful because its factors cancel exactly
in the standard parametrization. A complementary
quantity is obtained from a global unitary CKM fit,
\begin{equation}
\epsilon_V(\mu)\equiv
\frac{|V_{ts}(\mu)|}{|V_{cd}(\mu)|},
\label{eq:EpsilonVDefinition}
\end{equation}
where $\mu$ denotes the renormalization scale at which
the CKM matrix is evaluated. This scale dependence
becomes relevant when the experimentally determined
low-energy parameters are evolved towards a possible
flavour-matching scale.

Using the UTfit Summer-2025 Standard Model fit
\cite{cite:UTfit2025}, which combines available
flavour measurements within the three-generation
unitary CKM framework, we take
$|V_{ts}|=0.04117\pm0.00036$ and
$|V_{cd}|=0.22492\pm0.00062$. Identifying these
low-energy fit values with the boundary inputs
at the reference scale $\mu=m_t$ gives $\epsilon_V(m_t)\simeq0.18304\pm0.00168$.
The choice $\mu=m_t$ specifies a convenient
matching convention rather than a separate
experimental determination at that scale.
This result is close to the phenomenological
value in Eq.~\eqref{eq:EpsilonAdopted}, but
$\epsilon_V$ is defined through CKM moduli and
is not exactly equivalent to a ratio of
mixing angles.

The fitted moduli retain their dependence on the
physical CKM phase, without imposing a geometrical
ansatz for $\delta_{\rm CKM}$. Although $\epsilon_V$
is rephasing invariant, it is not phase independent.
Its  value is also
systematically below the geometrical hierarchy
parameter $\epsilon_C\simeq0.1869$ by about $2\%$.
This suggests a scale-dependent interpretation.
In the Wolfenstein expansion, $|V_{cd}|\sim\lambda$
and $|V_{ts}|\sim A\lambda^2$, hence
$\epsilon_V\simeq A\lambda$.
Under SM renormalization-group evolution, $\lambda$
is nearly scale independent whereas the parameters
controlling the $2$--$3$ and $1$--$3$ mixings
increase slowly with increasing
scale~\cite{cite:Antusch2026Running}.
The sign of the running is therefore precisely
the one required to evolve $\epsilon_V$
upward towards $\epsilon_C$.

The published SM running supports the possibility of a crossing
in the TeV range,
\begin{equation}
\epsilon_V(\Lambda_F)=\tan\alpha_C.
\label{eq:FNMatchingCrossing}
\end{equation}
The benchmark evolution in Ref.~\cite{cite:Antusch2026Running}
uses UTfit~2023 CKM inputs at $M_Z$, whereas the boundary inputs
adopted here are from UTfit~2025 and are assigned to $\mu=m_t$.
An approximate interpolation of the published running, normalised
to the adopted boundary inputs, suggests a crossing at a scale of
a few TeV. This estimate remains indicative pending a dedicated
RG evolution with consistent boundary conditions and matching.
The robust observation is that the required
$\simeq2\%$ shift is small and has the correct RG sign; the
present CKM uncertainties and matching conventions permit a broad
range of crossing scales.

This observation suggests interpreting the geometrical
value as a boundary condition on the flavour theory
rather than as an infrared equality,
\begin{equation}
\underbrace{\epsFN(\Lambda_F)=\epsilon_V(\Lambda_F)=\epsC}_{\mathrm{matching\ ansatz}},
\qquad \epsilon_V(m_t)\simeq0.183.
\label{eq:FNBoundaryFlow}
\end{equation}
Here the RG evolution connects $\epsilon_V(\Lambda_F)$ with $\epsilon_V(m_t)$; identifying $\epsilon_V(\Lambda_F)$ with the microscopic FN spurion is a separate matching assumption. In the usual FN notation the hierarchy parameter is given by
\begin{equation}
\epsFN=
\frac{\langle\Phi\rangle}{\Lambda_{\rm FN}},
\label{eq:FNSpurionScale}
\end{equation}
where $\Phi$ is the flavon field and
$\Lambda_{\rm FN}$ denotes the heavy messenger
or matching scale. If the crossing scale in
Eq.~\eqref{eq:FNMatchingCrossing} is tentatively
identified with $\Lambda_{\rm FN}$, the matching ansatz gives
\begin{equation}
\langle\Phi\rangle
\simeq\epsilon_C\Lambda_{\rm FN}
\simeq0.187\,\Lambda_{\rm FN}.
\label{eq:FNScaleEstimate}
\end{equation}
No numerical flavon vacuum expectation value is fixed without
a determination of $\Lambda_{\rm FN}$. Moreover, a generic low
FN scale may be strongly constrained by flavour-changing processes.
The role of this conditional relation is to exhibit a possible
physical meaning of the difference between the
low-energy hierarchy parameter near $0.183$
and the geometrical $P17$ value $\simeq0.1869$.

A decisive future check would be to evolve
complementary FN-sensitive quantities, including
CKM-modulus ratios and GST/mass combinations,
to a common scale. If they approach $\epsilon_C$ near
the same $\Lambda_F$, the interpretation in
Eq.~\eqref{eq:FNBoundaryFlow} would be
substantially stronger than the single crossing
in Eq.~\eqref{eq:FNMatchingCrossing}.

\section{Quark-mass hierarchy: cross-sector universality test}
\label{sec:QuarkMasses}

This section tests the second primary module, $H_{\rm tr}$:
whether the quadratic and bilinear hierarchy maps identified
in the mixing sector also describe the six quark masses,
with the bilinear scale specified as $m_F/\phiG$.
The relations are empirical correspondences among standard
reference parameters, not common-scale mass identities.
We use the PDG~2026 values~\cite{cite:PDG2026QuarkMasses}:
$\overline m_u(2~\mathrm{GeV})=2.16\pm0.07~\mathrm{MeV}$,
$\overline m_d(2~\mathrm{GeV})=4.70\pm0.07~\mathrm{MeV}$,
$\overline m_s(2~\mathrm{GeV})=92.9\pm0.7~\mathrm{MeV}$,
$\overline m_c(m_c)=1.2729\pm0.0045~\mathrm{GeV}$,
$\overline m_b(m_b)=4.186\pm0.006~\mathrm{GeV}$,
and $m_t^{(\mathrm{direct})}=172.60\pm0.27~\mathrm{GeV}$.
The quoted PDG intervals for the first five masses have 90\% confidence;
for the derived uncertainties below, we divide these intervals by
$1.645$ to obtain approximate Gaussian $1\sigma$ errors.
The direct top-mass uncertainty is taken as $1\sigma$.
The first five are $\overline{\mathrm{MS}}$ masses,
whereas the last is the directly reconstructed top-mass
parameter. These reference conventions form part of the
phenomenological comparison; no invariance under
renormalisation-group evolution or mass-scheme conversion
is assumed. Further qualifications are given in
Sec.~\ref{sec:RG}.

\subsection{Transfer hypothesis and the crossed mass ordering}
\label{subsec:MassOrdering}

In the quark-mixing hierarchy, the smallest angle is
generated by the quadratic map
$\mathcal Q(X,Y)=\sqrt{X^2+Y^2}-X
\simeq Y^2/(2X)$ for $Y\ll X$.
Applying this map to the two heaviest up-type masses gives
\begin{equation}
\mathcal Q(m_t,m_c)
\simeq\frac{m_c^2}{2m_t} =
4.694\pm0.021~\mathrm{MeV},
\label{eq:EmpiricalCrossingDiscovery}
\end{equation}
which agrees with the $d$-quark mass $\overline m_d(2~\mathrm{GeV})$  rather than the $u$-quark mass. 

The complementary bilinear map, $\mathcal B(X,Y;S)=XY/S$, applied to the remaining $b$,  $s$  and $u$ masses, gives an empirical scale
\begin{equation}
S_{\rm emp}\equiv\frac{m_bm_s}{m_u}
\simeq S,\qquad S\equiv\frac{m_F}{\phiG},
\label{eq:EmpiricalBSUScale}
\end{equation}
which is the larger fraction of the golden partition of $m_F$.
The two comparisons therefore motivate considering  the complementary
triples
\begin{equation}
(t,c,d),\qquad(b,s,u).
\label{eq:MassOrderedTriples}
\end{equation}

We call this arrangement \emph{crossed} because it differs
from the conventional SM charge-based ordering $(t,c,u)$
and $(b,s,d)$ by a single first-generation transposition
$u\leftrightarrow d$, yielding  the
heavier member of each generation in one row and the
lighter member in the other.  The hierarchy-map
comparisons and this heavy/light ordering criterion
therefore identify the same arrangement, but should not
be counted as statistically independent evidence.

The crossing concerns only the phenomenological
organisation of mass eigenvalues; it neither changes
their electroweak quantum numbers nor introduces a new
weak-interaction basis. Its dynamical origin remains
unexplained. The subsequent GST and FN interpretation
is likewise distinct from the empirical identification
of the crossed triples. In particular, the conventional
same-sector ratios $\sqrt{m_d/m_s}$ and $\sqrt{m_u/m_c}$
become cross-row ratios under this arrangement, providing
a further structural correspondence rather than a
derivation of the crossing.

\subsection{The $(t,c,d)$ Pythagorean hierarchy}
\label{subsec:TCD}

Applying the quadratic map of
Eq.~\eqref{eq:QuarkPythagorean} to the crossed triple
$(t,c,d)$ gives
\begin{equation}
m_d\simeq\sqrt{m_t^2+m_c^2}-m_t
\simeq\frac{m_c^2}{2m_t},
\qquad(m_c\ll m_t).
\label{eq:MassPythagoreanExact}
\end{equation}
The adopted reference masses give
$m_d^{(\mathrm{pred})}=4.694\pm0.021~\mathrm{MeV}$,
consistent with $m_d=4.70\pm0.07~\mathrm{MeV}$.

This relation has the algebraic structure of a two-state
seesaw and admits a Schur-complement interpretation
\cite{cite:SchurComplement}. Consider the illustrative
matrix
\begin{equation}
M_{\mathrm{toy}}=
\begin{pmatrix}
0&m_c\\
m_c&2m_t
\end{pmatrix}.
\label{eq:EffectiveSeesawBlock}
\end{equation}
Eliminating its heavy block yields the Schur complement
$-m_c^2/(2m_t)$, while the magnitude of its exact
light eigenvalue is $\sqrt{m_t^2+m_c^2}-m_t$.
This construction illustrates the algebraic relation
but does not constitute a gauge-invariant quark mass
model. A microscopic realisation connecting the
different electroweak sectors would require
additional structure.

\subsection{The $(b,s,u)$ bilinear hierarchy}
\label{subsec:BSU}

The complementary crossed triple $(b,s,u)$
exhibits the same product-over-scale structure
as the lepton bilinear relation~\eqref{eq:LeptonBilinear}:
\begin{equation}
m_u\simeq\frac{m_bm_s}{\mu_G},
\label{eq:MuBilinear}
\end{equation}
where  we have adopted $\mu_G\equiv m_F/\phiG$ as  the specified electroweak/golden normalisation scale, rather than an
adjustable mass parameter. The adopted reference inputs give $m_u^{(\mathrm{pred})}= 2.149\pm0.010~\mathrm{MeV}$, consistent with
$\overline m_u(2~\mathrm{GeV})$.

The two bilinear relations share the feature that the smallest quantity
is generated by a product of two larger quantities divided by a specified scale. 
This structural parallel is the content of the proposed cross-sector transfer,
rather than evidence for an established common dynamical mechanism.

At the reference conventions stated above, the two mass tests can
be summarised by the dimensionless closure ratios
\begin{equation}
\begin{aligned}
\mathcal C_Q&\equiv\frac{2m_tm_d}{m_c^2}
             =1.001\pm0.010,\\
\mathcal C_B&\equiv\frac{m_Fm_u}{\phiG m_bm_s}
             =1.005\pm0.020.
\end{aligned}
\label{eq:MassClosureDefinitions}
\end{equation}
Unity is the target of the leading quadratic and bilinear
correspondences, respectively. These ratios provide compact tests
of the same mass hypotheses, not additional independent evidence
or discovery significances.

\subsection{Effective indices and exploratory integer identifications}
\label{subsec:425}
Define
\begin{equation}
n_m\equiv\frac{\pi}{\arctan(m_c/m_t)},
\qquad
N_b\equiv\pi\frac{m_F/\phiG}{m_b}.
\label{eq:EffectiveMassIndices}
\end{equation}
Then $m_c/m_t=\tan(\pi/n_m)$ and $m_b/(m_F/\phiG)=\pi/N_b$ identically.  No integrality or literal polygonal interpretation is required by these definitions.  The adopted reference inputs give $n_m=426.0\pm1.1$ and $N_b=135.81\pm0.12$, consistent with $425=5^2\cdot17$ and $136=2^3\cdot17$, respectively.   Their numerical proximity to these integers is recorded as an exploratory arithmetic observation, identified after examining the mass ratios. The large effective indices may be regarded primarily as parametrisations of the observed mass hierarchies at the specified reference conventions. 

For the charm-to-top ratio, this gives the optional representation
\begin{equation}
\frac{m_c}{m_t}\simeq\tan\frac{\pi}{425}\simeq\frac{\pi}{425}.
\label{eq:McMt425}
\end{equation}
The doubling of the index follows from the Pythagorean map, not from an independent integer selection.  Writing $\widehat m_d=\sqrt{m_t^2+m_c^2}-m_t$ for the value supplied by that map gives the exact identity
$\widehat m_d/m_c=\tan\big(\frac12\arctan(m_c/m_t)\big) =\tan(\pi/2n_m)$.
Thus, conditional on the empirical mass correspondence and the exploratory assignment $n_m\simeq425$,
\begin{equation}
\frac{m_d}{m_c}\simeq\frac12\frac{m_c}{m_t}\simeq\frac{\pi}{850}.
\label{eq:MdMc850}
\end{equation}

In the bilinear branch, the analogous exploratory representation is
\begin{equation}
\frac{m_b}{m_F/\phiG}\simeq\frac{\pi}{136}
=\frac{\pi}{2^3\cdot17}\simeq0.02310.
\label{eq:Bottom136}
\end{equation}
Combining it with Eq.~\eqref{eq:MuBilinear} gives
\begin{equation}
\frac{m_u}{m_s}\simeq\frac{m_b}{m_F/\phiG}\simeq\frac{\pi}{136}.
\label{eq:MuMs136}
\end{equation}
The reference ratio $m_u/m_s=0.02325\pm0.00047$ is consistent with this value.  The index $850=2\cdot425$ and the second appearance of $136$ are dependent forms of the preceding relations, not additional independent evidence.  Although $136$ is a constructible polygon order and $425,850$ are not, no literal mass-sector polygons are assumed.

The recurrence of the same Fermat-prime factors already present in the mixing construction raises the question of whether an underlying selection rule or dynamics determines these combinations and their exponents.  No such rule is established here.  The principal empirical mass relations do not depend on the particular integer identifications, which may be tested separately as  the reference-parameter determinations improve.

\subsection{Summary of the transferred hierarchy maps}
\label{subsec:DualMaps}

Conditional on the transfer hypothesis, the mixing and mass sectors are summarized by the same two maps.  The first is quadratic,
$\mathcal{Q}(X_1,X_2) =\sqrt{X_1^2+X_2^2}-X_1 \simeq X_2^2/(2X_1)$,
while the second is bilinear,
$\mathcal{B}(X_1,X_2;S)=X_1X_2/S$.
The empirical assignments are
\begin{align}
\theta_{13}^q
&\simeq\mathcal{Q}(\theta_C,\theta_{23}^q),
&
 m_d
&\simeq\mathcal{Q}(m_t,m_c),
\label{eq:QuadraticAssignments}
\\
\theta_{13}^{\ell}
&\simeq\mathcal{B}(\theta_{12}^{\ell},\theta_{23}^{\ell};\pi),
&
 m_u
&\simeq\mathcal{B}(m_b,m_s;m_F/\phiG).
\label{eq:BilinearAssignments}
\end{align}
In the hierarchical limit, both maps exhibit an induced-coupling structure of the form ``product divided by a large scale''.  The leading quadratic expression is the symmetric case in which the two couplings are equal, whereas the bilinear map contains two different couplings and an external scale.  We refer to Eqs.~\eqref{eq:QuadraticAssignments} and~\eqref{eq:BilinearAssignments} collectively as dual induced-hierarchy relations.

Using published quark-mass determinations from the Fermilab Lattice, MILC and TUMQCD collaborations and the ETM collaboration~\cite{cite:Bazavov2018Masses,cite:Carrasco2014Masses,cite:Bussone2016Bottom}, we have also re-evaluated the two principal reference-scale mass relations while retaining the same top-mass input and electroweak normalisation.  The results remain numerically consistent with the proposed relations (robustness check against the choice of mass inputs).

\section{GST and FN interpretation of the primary hierarchy}
\label{sec:GSTFN}

\subsection{Standard GST relation}
\label{subsec:GST}

For a hierarchical symmetric $2\times2$ mass block
\begin{equation}
M=
\begin{pmatrix}
0&A\\
A&B
\end{pmatrix},
\qquad B\gg A,
\label{eq:GSTMatrix}
\end{equation}
with signed eigenvalues $-m_1$ and $m_2$,
diagonalisation gives $\tan^2\theta=m_1/m_2$.
For small mixing, this leads to the familiar GST
approximation $\theta_C\simeq\sqrt{m_d/m_s}$,
provided the down-sector rotation dominates.

The texture zero in Eq.~\eqref{eq:GSTMatrix} is not
required by the SM. For the general symmetric block
$M=\begin{pmatrix}D&A\\A&B\end{pmatrix}$,
with the same signed eigenvalues, one obtains
$\tan^2\theta=(m_1+D)/(m_2-D)$.
Thus even $D\ll m_2$ can appreciably affect the
mixing when $D$ is comparable to $m_1$.
For illustration, using the reference values of quark masses yields
$D\simeq0.25~\mathrm{MeV}$ in a pure down-sector two-family model. This is not a prediction, since
the physical Cabibbo angle generally involves rotations from both quark sectors.

Considering  the pair of GST-type correspondences  one obtains 
\begin{equation}
\theta_{12}^q\simeq\sqrt{\frac{m_d}{m_s}}
=0.2249\pm0.0011,\qquad
\theta_{23}^q\simeq\sqrt{\frac{m_u}{m_c}}
=0.04119\pm0.00041,
\label{eq:SequentialGST}
\end{equation}
The Cabibbo and $2$--$3$ correspondences differ from the global-fit
angles by approximately $1.6\sigma$ and $0.8\sigma$, respectively,
when the quoted uncertainties are combined in quadrature.
The second correspondence is not part of the conventional GST assignment; together, we refer
to Eq.~\eqref{eq:SequentialGST} as the crossed or sequential GST approximation.

Substituting these relations into the geometrical hierarchy of Eq.~\eqref{eq:QuarkPythagorean} gives
\begin{equation}
\sqrt{\frac{m_d}{m_s}+\frac{m_u}{m_c}}
-\sqrt{\frac{m_d}{m_s}}
=0.003741\pm0.000076
=(0.2143\pm0.0043)^\circ.
\label{eq:GSTMassTriangle}
\end{equation}
The result agrees with the measured $\theta_{13}^q$ within the present uncertainties. This correspondence
expresses the same Pythagorean hierarchy in terms of quark-mass ratios, without requiring exact texture zeros
or introducing an additional independent mass relation.

\subsection{FN charge interpretation of the crossed relation}
\label{subsec:FNCrossed}

The correspondence $\theta_{23}^q\simeq\sqrt{m_u/m_c}$
does not identify the CKM $2$--$3$ angle with an intrinsic
up-sector $1$--$2$ rotation. Together with
$\theta_C\simeq\sqrt{m_d/m_s}$, it motivates a comparison
of hierarchy powers. The FN interpretation below concerns
compatibility with Yukawa power counting, not a derivation
of the crossed mass assignment in Sec.~\ref{sec:QuarkMasses}.

For a single FN spurion $\epsFN=\langle\Phi\rangle/\Lambda$
and an uncharged Higgs~\cite{cite:FN}, let $Q_i^{(q)}$,
$Q_i^{(u)}$ and $Q_i^{(d)}$ denote the FN charges
of the left-handed quark doublets, up-type singlets and
down-type singlets, respectively. We use the additive
convention $(Y_u)_{ij}\sim\epsFN^{Q_i^{(q)}+Q_j^{(u)}}$ and
$(Y_d)_{ij}\sim\epsFN^{Q_i^{(q)}+Q_j^{(d)}}$. With generic order-one
coefficients and no cancellations altering the leading powers,
the left-handed mixing scales as
$\theta_{ij}^q\sim\epsFN^{|Q_i^{(q)}-Q_j^{(q)}|}$.
The hierarchy $\theta_C\sim\epsFN$,
$\theta_{23}^q\sim\epsFN^2$ and
$\theta_{13}^q\sim\epsFN^3$ therefore requires
$Q_1^{(q)}-Q_2^{(q)}=1$ and $Q_2^{(q)}-Q_3^{(q)}=2$ for ordered charges.
Choosing $Q_3^{(q)}=0$ gives
\begin{equation}
Q^{(q)}=(3,2,0).
\label{eq:FNLeftCharges}
\end{equation}
For ordered singlet charges, the mass powers are
$m_{u_i}\sim\epsFN^{Q_i^{(q)}+Q_i^{(u)}}$ and
$m_{d_i}\sim\epsFN^{Q_i^{(q)}+Q_i^{(d)}}$.
The GST-type scalings then require $Q_1^{(u)}-Q_2^{(u)}=3$ and
$Q_1^{(d)}-Q_2^{(d)}=1$. An unsuppressed top Yukawa fixes $Q_3^{(u)}=0$;
the illustrative choices $Q_2^{(u)}=1$ and $Q_2^{(d)}=Q_3^{(d)}=2$ give
\begin{equation}
Q^{(u)}=(4,1,0),\qquad Q^{(d)}=(3,2,2).
\label{eq:FNRightCharges}
\end{equation}
The corresponding textures are
\begin{equation}
Y_u\sim
\begin{pmatrix}
\epsFN^7&\epsFN^4&\epsFN^3\\
\epsFN^6&\epsFN^3&\epsFN^2\\
\epsFN^4&\epsFN&1
\end{pmatrix},\qquad
Y_d\sim
\begin{pmatrix}
\epsFN^6&\epsFN^5&\epsFN^5\\
\epsFN^5&\epsFN^4&\epsFN^4\\
\epsFN^3&\epsFN^2&\epsFN^2
\end{pmatrix},
\label{eq:FNYukawaTextures}
\end{equation}
where each entry carries its own order-one coefficient.
Their singular values yield
\begin{equation}
m_u:m_c:m_t\sim\epsFN^7:\epsFN^3:1,\qquad
m_d:m_s:m_b\sim\epsFN^6:\epsFN^4:\epsFN^2.
\label{eq:FNHiearchies}
\end{equation}
These powers reproduce both GST-type scalings but do not
fix their numerical coefficients relative to the CKM angles.

The same CKM and GST power conditions admit the broader
ordered, non-negative integer charge family
\begin{equation}
\begin{gathered}
Q^{(q)}=(3,2,0),\quad Q^{(u)}=(p+3,p,0),\\
Q^{(d)}=(r+s+1,r+s,r),\qquad p,r,s\in\mathbb Z_{\geq0},
\end{gathered}
\label{eq:FNChargeFamily}
\end{equation}
with mass exponents
\begin{equation}
(n_u,n_c,n_t)=(p+6,p+2,0),\quad
(n_d,n_s,n_b)=(r+s+4,r+s+2,r).
\label{eq:FNFamilyMassPowers}
\end{equation}
Thus $n_u-n_c=4$ and $n_d-n_s=2$ throughout; the
representative above is $(p,r,s)=(1,2,0)$.
Before fixing the charge normalisation, the relabelling
$Q_i^{(q)}\mapsto Q_i^{(q)}+t$, $Q_i^{(u)}\mapsto Q_i^{(u)}-t$ and
$Q_i^{(d)}\mapsto Q_i^{(d)}-t$ leaves all Yukawa exponents unchanged,
but need not relate equivalent ultraviolet theories.

For the numerical comparison we use $\epsql$ from
Eq.~\eqref{eq:EpsilonAdopted} as a phenomenological expansion
parameter. Identifying it with the microscopic spurion
$\epsFN$ is an additional assumption, since mixing-angle
ratios also contain order-one coefficients.
For the chosen representative, define
\begin{align}
m_u&=\cF{u}m_F\epsql^7,&
 m_c&=\cF{c}m_F\epsql^3,&
 m_t&=\cF{t}m_F,\nonumber\\
m_d&=\cF{d}m_F\epsql^6,&
 m_s&=\cF{s}m_F\epsql^4,&
 m_b&=\cF{b}m_F\epsql^2.
\label{eq:FNCoefficientDefinition}
\end{align}
The $\cF{i}$ are effective coefficients of the specified
reference masses, not individual Yukawa-matrix entries.
This representative matches the quadratic mass law in
powers, $n_d=2n_c-n_t=6$, whereas the bilinear law has
$n_u=7\ne n_b+n_s=6$. The two empirical mass relations
therefore require, respectively,
$\cF{d}\simeq\cF{c}^{\,2}/(2\cF{t})$ and
$\epsql\cF{u}\simeq\phiG\cF{b}\cF{s}$;
neither coefficient condition follows from the FN charges.

Defining $\kappa_u^2=\cF{u}/\cF{c}$ and
$\kappa_d^2=\cF{d}/\cF{s}$ gives
\begin{equation}
\sqrt{m_u/m_c}=\kappa_u\epsql^2,\qquad
\sqrt{m_d/m_s}=\kappa_d\epsql,
\label{eq:GSTFNWithKappa}
\end{equation}
and hence
\begin{equation}
\frac{m_um_s}{m_cm_d}={\cal C}_F\epsql^2,\qquad
{\cal C}_F\equiv\frac{\cF{u}\cF{s}}{\cF{d}\cF{c}}
=\frac{\kappa_u^2}{\kappa_d^2}.
\label{eq:EpsilonMassRelation}
\end{equation}
Since $n_u+n_s-n_d-n_c=2$ throughout
Eq.~\eqref{eq:FNChargeFamily}, the value of ${\cal C}_F$
at fixed masses and $\epsql$ is independent of the charge
representative.

For the numerical estimates below, ratio errors retain the shared
dependence on $\epsql$, using the $1\sigma$ mass-error convention
specified in Sec.~\ref{sec:QuarkMasses}.
With the adopted reference masses and
$\epsql=0.1833\pm0.0028$, the coefficients are
$(\cF{u},\cF{c},\cF{t})\simeq
(1.062\pm0.115,\,0.706\pm0.032,\,0.5895\pm0.0009)$
and $(\cF{d},\cF{s},\cF{b})\simeq
(0.423\pm0.039,\,0.281\pm0.017,\,0.426\pm0.013)$,
all of order unity.
In particular, $\cF{u}/\cF{c}=1.504\pm0.096$ and
$\cF{d}/\cF{s}=1.506\pm0.048$, while
$\cF{u}/\cF{d}=2.507\pm0.066$ and
$\cF{c}/\cF{s}=2.511\pm0.040$.
Both pairwise equalities express the same condition,
$\cF{u}\cF{s}\simeq\cF{d}\cF{c}$, quantified by
\begin{equation}
{\cal C}_F=0.998\pm0.038.
\label{eq:FNCoefficientCrossRatio}
\end{equation}
No significance is assigned to the proximity of the ratios
to $3/2$ or $5/2$. Their near-factorisation concerns an
auxiliary array of effective mass coefficients, not the
rank of either physical Yukawa matrix.

Replacing $\epsql$ by the fixed geometrical value
$\epsC=\tan\alpha_C$ gives
\begin{equation}
{\cal C}_F^{(\mathrm{P17})}
\equiv\frac{m_um_s}{m_cm_d\epsC^2}
=0.960\pm0.021,
\label{eq:Epsilon17CoefficientTest}
\end{equation}
about $1.9\sigma$ below unity under this approximation.
The two cross-ratio evaluations use the same masses and
are not independent tests. Their possible flavour-scale
interpretation requires controlled running and matching,
as discussed in Secs.~\ref{subsec:FNMatchingScale}
and~\ref{sec:RG}. The FN construction thus organises the
hierarchy powers, while the observed coefficient relations
remain empirical rather than consequences of a derived
flavour mechanism.

\section{Reference scales, running, and FN coefficients}
\label{sec:RG}

The mass relations use different, standard reference conventions:
$\overline m_{u,d,s}(2~\mathrm{GeV})$,
$\overline m_c(m_c)$ and $\overline m_b(m_b)$ in the
$\overline{\mathrm{MS}}$ scheme, together with
$m_t^{(\mathrm{direct})}$ from kinematic reconstruction
\cite{cite:PDG2026QuarkMasses,cite:PDG2026Top}.
The light-quark and heavy-quark self-scales are conventional
QCD reference points, whereas the directly reconstructed
top parameter is not itself a specified short-distance
running mass. Below the top threshold the top is integrated
out, and converting its direct mass to a running parameter
requires an explicit prescription and threshold matching.
The two reference-scale correspondences are therefore
\begin{equation}
\overline m_d(2~\mathrm{GeV}) \simeq\frac{[\overline m_c(m_c)]^2}{2m_t^{(\mathrm{direct})}}, \qquad
\overline m_u(2~\mathrm{GeV}) \simeq\, \frac{\overline m_b(m_b)\,\overline m_s(2~\mathrm{GeV})}{m_F/\phiG}.
\label{eq:RGReferenceRelations}
\end{equation}
Their simple coefficients are not asserted to be
renormalisation-group or scheme invariant.

There is nevertheless limited stability under low-energy
running. In pure QCD, quark masses in the same active-flavour
theory have a common anomalous dimension; the ratios
$m_d/m_s$ and $m_u/m_s$ at a shared scale are therefore
unchanged by this running, up to smaller non-QCD effects.
Accordingly, $m_u(\mu)/m_s(\mu)$ can be compared with the
fixed reference quantity $m_b(m_b)/(m_F/\phiG)$ without
altering the leading pure-QCD running of the light-mass
ratio. Likewise, $m_c^2/m_t$ and $m_d$ would carry the same
net QCD mass-renormalisation factor if all were running
masses in one active-flavour theory. Neither observation
removes the distinct reference conventions and top-threshold
matching required by Eq.~\eqref{eq:RGReferenceRelations}.

At a common flavour scale $\Lambda_F$, FN coefficients are
defined through running Yukawa eigenvalues~\cite{cite:Antusch2026Running}, schematically
$m_i(\Lambda_F)=\cF{i}(\Lambda_F)m_F[\epsFN(\Lambda_F)]^{n_i}$.
The corresponding cross ratio ${\cal C}_F(\mu)$ need not
remain unity under evolution. Writing
$\cF{i}^{(\mathrm{ref})}=Z_i\cF{i}(\Lambda_F)$, where
$Z_i$ includes the required running and matching factors,
gives
\begin{equation}
{\cal C}_F^{(\mathrm{ref})}
={\cal C}_F(\Lambda_F)\frac{Z_uZ_s}{Z_dZ_c}.
\label{eq:CrossRatioMatching}
\end{equation}
The observed ${\cal C}_F^{(\mathrm{ref})}\simeq1$ is therefore
a condition on this product, not a prediction of generic
Abelian FN charges or an RG-invariant identity.
A microscopic theory would need to calculate the coefficient
relations, the threshold factors and the conversion of
$m_t^{(\mathrm{direct})}$ to a suitable mass convention. Until then,
Eq.~\eqref{eq:RGReferenceRelations} remains a pair of
empirical relations at explicitly specified reference points.

\section{The Fermi scale and quark--lepton mass correspondences}
\label{sec:TopFermi}

The golden constructions suggest a further empirical cross-sector
correspondence. With $m_F=G_F^{-1/2}$ as defined in
Sec.~\ref{sec:Introduction}, the adopted inputs give
\begin{equation}
f_t^{(\mathrm{exp})}\equiv\frac{m_t^{(\mathrm{direct})}}{m_F}
=0.589468\pm0.000922.
\label{eq:ftTopDefinition}
\end{equation}
The two prescriptions give
$f_t^{(P5)}\equiv\cos\alpha_{P5}\simeq0.587785$ and
$f_t^{(G)}\equiv\tfrac12\arccos(\phiG^{-2})\simeq0.589437$.
The $G$ value agrees particularly closely with the measured
ratio, whereas the $P5$ value lies about $1.8\sigma$ below it.
We retain both as alternative phenomenological identifications.

Combining $m_t=m_F f_t$ with the leading approximation
$m_c/m_t\simeq\pi/425$ in Eq.~\eqref{eq:McMt425} and
the quadratic mass relation~\eqref{eq:MassPythagoreanExact}
gives $m_d\simeq m_F f_t\pi^2/(2\cdot425^2)$.
For either branch, this defines an inferred Fermi constant,
\begin{equation}
G_F^{(P5,G)}\equiv
\frac{\left(f_t^{(P5,G)}\right)^2}{4m_d^2}
\left(\frac{\pi}{425}\right)^4.
\label{eq:GFmu}
\end{equation}
Using $m_d\equiv\overline m_d(2~\mathrm{GeV})
=4.70~\mathrm{MeV}$~\cite{cite:PDG2026QuarkMasses}
and the adopted mass-error convention yields the approximate
$1\sigma$ results
$G_F^{(P5)}\simeq(1.167\pm0.021)\times10^{-5}
~\mathrm{GeV}^{-2}$ and
$G_F^{(G)}\simeq(1.174\pm0.021)\times10^{-5}
~\mathrm{GeV}^{-2}$.
Their unrounded central values exceed the muon-decay
determination
$G_F=1.16637859(59)\cdot 10^{-5}~\mathrm{GeV}^{-2}$
\cite{cite:FermiConstantUpdate2026}
by approximately $0.089\%$ and $0.652\%$, respectively.
Both are compatible with this determination; the propagated
uncertainties, with $f_t^{(P5,G)}$ and the index $425$ fixed,
are about $1.8\%$. Thus $G$ gives the closer top--Fermi
ratio, while $P5$ gives the closer central value in this
combined mass comparison, which does not discriminate
between the branches at present precision.

This reformulation connects the down-quark reference mass
to the weak-interaction scale through golden algebra and
the exploratory index $425$. It is a conditional consequence
of the preceding mass hypotheses, not an independent
derivation of $G_F$. Both this relation and the top--Fermi
correspondence depend on the specified mass conventions;
a microscopic explanation would also have to account
for running and matching.

Although charged-lepton masses are not part of the primary
hierarchy-transfer hypothesis, one further empirical correspondence
may be relevant to a more complete description of fermion masses.
We use the PDG~2026 charged-lepton pole masses
$m_\mu=(105.6583755\pm0.0000023)~\mathrm{MeV}$ and
$m_\tau=(1776.93\pm0.09)~\mathrm{MeV}$~\cite{cite:PDG2026}.
Applying the leading quadratic, seesaw-like expression to the
$(\tau,\mu)$ pair gives
$m_\mu^2/(2m_\tau)=(3.14129\pm0.00016)~\mathrm{MeV}$,
rather than the electron mass, $m_e\simeq0.511~\mathrm{MeV}$.
An exploratory dimensionless correspondence is
$m_\mu^2/(2m_\tau m_F)\simeq\pi^3/(16\cdot425^2)$,
which agrees with the adopted charged-lepton masses and Fermi
scale to about $5.6\times10^{-5}$ relatively.
Combining this near-relation with
$m_d\simeq m_F f_t^{(P5,G)}\pi^2/(2\cdot425^2)$,
as discussed above, eliminates the common hierarchy factor and
suggests
\begin{equation}
\frac{m_\mu^2}{2m_\tau}
\simeq\frac{\pi}{8f_t^{(P5,G)}}\,m_d.
\label{eq:MuTauDownCorrespondence}
\end{equation}
For the adopted reference value
$\overline m_d(2~\mathrm{GeV})=4.70~\mathrm{MeV}$,
the right-hand side, evaluated at the central input masses, is
$3.1401~\mathrm{MeV}$ for $P5$ and
$3.1313~\mathrm{MeV}$ for $G$. The relative central-value
differences from $m_\mu^2/(2m_\tau)$ are approximately
$3.9\times10^{-4}$ and $3.2\times10^{-3}$, respectively,
both well below the uncertainty inherited from $m_d$.
The relation links charged-lepton pole masses to a specified
running down-quark mass, and is exploratory rather than
renormalisation-group invariant. It is also algebraically connected
to the preceding $425$ hierarchy, not an independent test of its
integer assignment. A microscopic account would have to explain
both the common suppression factor and the required mass matching.

\section{Symmetry interpretation and outlook}
\label{sec:Symmetry}

The symmetry discussion is deliberately limited to compatibility statements and model-building targets.  The $P5$ solution can be embedded in the expected pentagonal $D_5/Z_5$ symmetry structure, but such an embedding is constructed around the geometrical solution and is not an independent derivation of its physical orientation.  In particular, a reflection chosen to leave the $P5$ vector invariant, or a fivefold angular potential chosen to possess the corresponding extrema, demonstrates that the geometry can be expressed in symmetry language; it does not show that a physical residual flavour symmetry predicts the observed mixing vector.  The exact content retained here is therefore the pentagonal/Chebyshev geometry itself, while the dynamical selection of the physical direction remains open.

The $G$ construction supplies a second exact auxiliary algebraic structure: the reciprocal roots $\phiG^{-2}$ and $\phiG^2$ occupy the circular and hyperbolic branches of the Chebyshev continuation, $T_n(\cos\alpha)=\cos(n\alpha)$ and
$T_n(\cosh\eta)=\cosh(n\eta)$, while the solar probability is the exact multiplicative midpoint between the GR1 and GR2 golden-ratio predictions.  Here the circular--hyperbolic language refers only to the analytic continuation of the same reciprocal algebraic input; no noncompact physical degree of freedom is being introduced.  Whether such an auxiliary structure has a dynamical realization in flavour space remains open.

Known golden-ratio mixing schemes show that fivefold structures can arise from genuine flavour symmetries; in particular, GR1 admits $A_5$-based realizations~\cite{cite:GoldenA5}.  The proximity of the $P5/G$ points to such symmetry limits therefore makes $A_5$ a useful reference framework rather than an identification of the symmetry underlying the present construction.  Residual $A_5$ and generalised-CP models can correlate the PMNS parameters while leaving one continuous parameter undetermined~\cite{cite:DiIuraA5CP}.  An explanation of the specific $P5/G$ values would additionally have to determine the family representations, flavon or spurion content, invariant Yukawa operators, vacuum alignments and any remaining continuous parameters.  The present geometric relations are consequently proposed as quantitative targets for a microscopic flavour theory, not as predictions of a completed finite-group model.

A complete theory would also have to connect this lepton structure to the $P17$ quark hierarchy, the crossed quark-mass organization, the FN coefficient correlations, and the appearance of the same golden algebra in the dimensionless top--Fermi ratio.  None of these ingredients is presently derived from a single Lagrangian.

The present study concerns mixing magnitudes and mass hierarchies and does not propose a mechanism for CP violation.  The experimentally fitted quark CP phase admits more than one approximate numerical representation using the geometrical ingredients considered here; such non-unique and related correspondences are neither imposed as constraints nor counted as additional evidence.  No CP-phase ansatz is required for the principal CKM-modulus ratios or the PMNS combinations in Eq.~\eqref{eq:PMNSAnglesFromModuli}, whose phase independence follows from the standard parametrizations~\cite{cite:PDGCKMParam,cite:PDGPMNSParam}.  In phase-sensitive auxiliary quantities, notably $\epsilon_V$, the experimentally fitted CKM information and its phase dependence are retained as described in Sec.~\ref{subsec:FNMatchingScale}.  The real geometric relations therefore neither imply CP conservation nor explain the origin or sign of CP violation~\cite{cite:Jarlskog}; these remain requirements for a microscopic completion.

The empirical reference-scale mass relations in Eq.~\eqref{eq:RGReferenceRelations} also suggest a possible connection between flavour hierarchies and nuclear viability.  Conditional on these relations, variations of $m_c$ and $m_s$, with $m_t$, $m_b$, $m_F$, the gauge couplings and the remaining reference inputs held fixed, induce corresponding variations of $m_u+m_d$ and $m_d-m_u$.  This provides a parametrisation for a future investigation, rather than an established nuclear constraint on the intermediate quark masses.  A physical assessment would require a consistent treatment of parameter variations, threshold effects, the QCD scale $\Lambda_{\mathrm{QCD}}$, and the relevant hadronic and nuclear calculations~\cite{cite:DamourDonoghue2008,cite:JaffeJenkinsKimchi2009,cite:Borsanyi2015Np}.

\section{Discussion}
\label{sec:Discussion}

\subsection{The $P5/G$ near-degeneracy: interpretation and limitations}
\label{subsec:P5GDoublet}

The near-degenerate $P5$ and $G$ prescriptions might be interpreted
as a dynamical doublet of flavour--mass alignments, rather than as
additional neutrino flavours. Stochastic mixing and scalar-induced
time-dependent mixing provide related precedents~\cite{cite:Guzzo2013Stochastic,cite:Krnjaic2018DiNO},
but neither derives this specific geometrical pair, for which no
established microscopic implementation is presently available.
Even distinguishing the two pure prescriptions would require
exceptionally high precision (Sec.~\ref{subsec:PrecisionProspects}).
Establishing coexistence of the two branches is a separate question,
which present oscillation data cannot resolve.
In particular, for a common mass spectrum and an unresolved mixture
with one branch applying throughout each event, two-flavour vacuum
disappearance is exactly equivalent to that produced by a single
effective mixing angle. More general observables could in principle
break this degeneracy, but no experimentally accessible signature
of the proposed doublet has been established. We therefore retain
$P5$ and $G$ as alternative phenomenological prescriptions, not as
evidence for physically split mixing angles.

\subsection{The unassigned first-level lepton angle as a sterile-mixing benchmark}

Alongside the empirical reactor-angle relation
$\theta_{13}^{\ell}\simeq\theta_{23}^{\ell}\epsilon_C$,
the $P5$--$P17$ hierarchy contains a companion quantity obtained
by applying the same first-level suppression to the solar angle,
\begin{equation}
\theta_x \equiv \theta_{12}^{\ell}\epsilon_C
\simeq 6.30^\circ.
\label{eq:SterileAnglePrediction}
\end{equation}
The numerical value here follows from the geometric $P5$ solar
angle and $\epsilon_C=\tan\alpha_C$. This quantity arises at the
same hierarchy level as the reactor angle, rather than from
a speculative continuation to higher powers of $\epsilon_C$.
There is no additional independent mixing angle in the
$3\times3$ PMNS parametrisation to which it can be assigned.
We therefore regard $\theta_x$ as an unassigned ``missing partner'',
without assuming which additional rotation, if any, it represents.

One possible interpretation arises in a minimal $3+1$ neutrino
scheme through the conjecture
\begin{equation}
\theta_x=\theta_{14}.
\label{eq:Theta14Identification}
\end{equation}
In the conventional $3+1$ parametrisation,
$|U_{e4}|^2=\sin^2\theta_{14}$. Replacing the geometric solar angle
and hierarchy parameter by their adopted experimental estimates
gives the conditional benchmark
$\theta_{14}=(6.17\pm0.17)^\circ$ and
$|U_{e4}|^2=0.0115\pm0.0006$.
These are inferred quantities under
Eq.~\eqref{eq:Theta14Identification}, not direct measurements
of sterile mixing.

In the standard vacuum short-baseline approximation, the
corresponding electron-antineutrino disappearance amplitude $\sin^2 2\theta_{14} =4|U_{e4}|^2(1-|U_{e4}|^2)$  amounts to
\begin{equation}
\sin^2 2\theta_{ee} =\sin^2 2\theta_{14} =0.0456\pm0.0024.
\label{eq:SterileAmplitudePrediction}
\end{equation}
Using the geometric $P5$--$P17$ values instead gives
$\sin^2 2\theta_{ee}\simeq0.0475$, with no additional continuous
parameter once the identification $\theta_x=\theta_{14}$ is imposed.
The mass-squared splitting $\Delta m_{41}^2$ remains unspecified
by the construction.

These amplitudes can be compared with published short-baseline
reactor analyses. The 2018 combined NEOS--DANSS spectral-ratio analysis,
which avoids reliance on an absolute reactor-flux prediction,
found a preferred island near $\Delta m_{41}^2\simeq1.3~{\rm eV}^2$ and  $\sin^2 2\theta_{ee}=0.049\pm0.023$~ $(2\sigma)$,
where the amplitudes are numerically compatible with both benchmarks
\cite{cite:NEOSDANSS2018}.
A later, largely reactor-model-independent RENO--NEOS analysis
found its global best fit at $|\Delta m_{41}^2|=(2.41\pm0.03)~{\rm eV}^2$ and 
$\sin^2 2\theta_{14}=0.08\pm0.03$\cite{cite:RENONEOS}.
The DANSS analysis cited here also reports no statistically
significant sterile-neutrino signal. It finds two local minima near
$(\Delta m_{41}^2,\sin^2 2\theta_{ee})
\simeq(0.3~{\rm eV}^2,0.07)$ and  $(\Delta m_{41}^2,\sin^2 2\theta_{ee}) \simeq(1.3~{\rm eV}^2,0.016)$,
with a significance of about $2\sigma$ for the preferred minimum\cite{cite:DANSSRecent}.
The second minimum lies near the older NEOS--DANSS oscillation
frequency but favours a smaller amplitude.
The final STEREO analysis and the final PROSPECT-I search find
no significant short-baseline oscillation signal and exclude
substantial regions of eV-scale sterile-neutrino parameter space
\cite{cite:STEREOFinal,cite:PROSPECTFinal}.

These reactor results do not establish a common, statistically
significant sterile-neutrino region associated with
Eq.~\eqref{eq:SterileAnglePrediction}. Some reported local best-fit
amplitudes are numerically comparable to the proposed value, but
this does not establish experimental support for the geometrical
identification. At present, the construction supplies only the
conditional benchmark $\theta_{14}=\theta_x\simeq6.30^\circ$,
corresponding to $|U_{e4}|^2\simeq0.0120$ and
$\sin^2 2\theta_{ee}\simeq0.0475$.
Its viability must be assessed as a function of the unspecified
$\Delta m_{41}^2$, including the null-search constraints;
agreement with an individual best-fit amplitude alone does not
establish compatibility with the combined reactor data.

\subsection{Prospects for sharper experimental tests}
\label{subsec:PrecisionProspects}

In the lepton sector, improved determinations of
$\theta_{23}^{\ell}$, particularly its octant, would test
the common $P5/G$ prediction near $46.3^\circ$;
a confirmed lower-octant value would disfavour both
constructions. More precise solar and reactor measurements
would further constrain the proposed mixing hierarchy,
although distinguishing the two fixed $P5/G$ predictions
would require exceptional precision. For example, their solar-angle
separation, $\Delta\theta_{12}^{\ell}\simeq0.0946^\circ$, requires
$\sigma(\theta_{12}^{\ell})\lesssim0.019^\circ$ for a nominal
$5\sigma$ discrimination, treating both prescriptions as exact.
Equivalently, $\sigma(\sin^2\theta_{12}^{\ell})\lesssim3.1\times10^{-4}$,
about $0.10\%$ relative precision, would be needed.

In the quark sector, improved and mutually consistent
determinations of $|V_{us}|$, $|V_{cb}|$ and $|V_{ub}|$
would test whether the present global-fit preference for the
golden-radius Cabibbo prescription persists and sharpen tests
of the $17\to34$ hierarchy.
In particular, the present inclusive--exclusive tension
in $|V_{cb}|$ must be resolved. A possible flavour-scale
interpretation would additionally require controlled
RG evolution and matching.

The transferred mass relations would benefit from
improved lattice determinations of light-quark masses,
together with consistent mass
definitions and top-mass matching. The decisive test is
whether both closure ratios, $\mathcal C_Q$ and $\mathcal C_B$
in Eq.~\eqref{eq:MassClosureDefinitions}, remain compatible with
unity as uncertainties decrease. The exploratory sterile-neutrino
interpretation could likewise be tested through
short-baseline disappearance measurements, conditional
on $\Delta m_{41}^2$. Throughout, the geometrical
prescriptions and hierarchy maps must remain fixed
so that new measurements provide genuine prospective tests.

\section{Conclusions}
\label{sec:Conclusions}

The empirical analysis supports a coherent phenomenological
pattern in which flavour-mixing magnitudes are organised,
directly or indirectly, by golden-ratio geometry and the
golden partition of unity. Its significance lies not in
isolated numerical agreements, but in the recurrence of
related structures across quark and lepton observables.
This motivates the hypothesis that golden/pentagonal
algebra provides a common organising principle for flavour
mixing, rather than a separate prescription for each angle.

Parametrising suppression ratios by effective polygonal
orders provides a useful way of comparing hierarchies
across sectors. The nontrivial test is whether the same
fixed orders describe several observables, not whether
each ratio can be assigned a nearby integer. The
fifth-order Chebyshev structure and the heptadecagonal
$17\to34$ hierarchy suggest a role for discrete orders,
including the Fermat primes $5$ and $17$ and simple
integer multiples. Their physical selection remains
unexplained; the larger mass-sector integer
identifications remain exploratory.

The lepton and quark suppression ratios are compatible
with a common phenomenological hierarchy parameter $\epsql$. Moreover,
the approximate relation
$\tan\alpha_C\simeq\cos54^\circ/\pi$ connects its
heptadecagonal realisation to the golden/pentagonal
large-angle structure. This suggests that angular
normalisation and hierarchical suppression may be
related aspects of flavour organisation. Such universality
is presently phenomenological: identifying the common
parameter with a microscopic FN spurion or a high-scale
boundary value requires additional dynamical and
matching assumptions.

Two forms of induced hierarchy are distinguished:
quadratic, seesaw-like suppression and bilinear
product-over-scale suppression. Their recurrence in
the crossed mass triples $(t,c,d)$ and $(b,s,u)$
motivates a further universality hypothesis linking
mixing and mass hierarchies. The crossing then organises
possible mass-generation relations rather than changing
quark quantum numbers. A microscopic explanation must
account for the functional forms and their normalisations,
including the running and matching needed to reproduce
the specified reference-mass correspondences.

The near-degeneracy of the $P5$ and $G$ prescriptions
and the approximate factorisation of the effective FN
coefficients suggest further relations that may reflect
residual or weakly broken flavour symmetries. These
observations provide quantitative targets for such
interpretations, but neither numerical proximity nor
algebraic factorisation establishes a symmetry of the
underlying theory.

The fixed alternatives also provide discriminating tests.
More precise and mutually consistent Cabibbo-angle
determinations can test the persistence of the present global-fit
preference for the golden-radius normalisation over the solar-derived
alternative, while the atmospheric octant tests a prediction common
to both $P5$ and $G$. Improved mass determinations test the
cross-sector transfer through $\mathcal C_Q$ and $\mathcal C_B$,
separately from the mixing hypothesis. The strongest support would come
from the persistence of the correlated pattern under
new measurements, without readjusting its defining
prescriptions.

The resulting framework is therefore a set of connected
phenomenological constraints for a microscopic flavour
theory, not a derived mechanism or an explanation of
CP violation. Its central open question is whether the
golden geometry, common suppression parameter and two
hierarchy maps follow from a single physical structure.

\begin{acknowledgments}
The core geometrical framework, empirical analysis  and physical hypotheses were developed by the author, who takes full responsibility for the scientific claims, calculations, references, and final text. Symbolic and numerical calculations were performed in part by the author using Mathematica~14.3 (Wolfram Research). ChatGPT-6 ASTRA (OpenAI) was used as an interactive research and writing aid, with assistance in literature searches, critical assessment of mathematical relations, cross-checking algebraic and numerical calculations, and improving language and readability. 
\end{acknowledgments}

\end{document}